\documentclass[pra,aps,nofootinbib,showpacs,floatfix,superscriptaddress,amsmath,twocolumn]{revtex4}
\usepackage{graphicx}
\usepackage{amsmath}
\usepackage{amssymb}
\usepackage{color}

\newcommand {\fabs}[1] {\left| #1 \right|}
\newcommand {\fexp} [1] {\exp \left( #1 \right)}
\newcommand {\fcos} [1] {\cos \left( #1 \right)}
\newcommand {\fsin} [1] {\sin \left( #1 \right)}

\newcommand {\la}{\langle}
\newcommand {\ra}{\rangle}

\newcommand{\ket}[1]{\ensuremath{|#1\rangle}}
\newcommand{\bra}[1]{\langle#1|}
\newcommand{\avg}[1]{\left \langle #1\right \rangle}
\newcommand{\braket}[2]{\langle#1|#2\rangle}
\newcommand{\ketbra}[2]{|#1\rangle\langle#2|}

\newcommand{\Sket}[1]{\ensuremath{|#1\rangle\!\rangle}}
\newcommand{\Sbra}[1]{\langle\!\langle#1|}
\newcommand{\Sbraket}[2]{\langle\!\langle#1|#2\rangle\!\rangle}

\newcommand {\bN}{\mathbb{N}}

\newcommand{\cL}{{\cal{L}}}

\newcommand{\cU}{{\cal{U}}}
\newcommand{\tr}{{\mbox{tr}}}

\begin{document}
\title{Effect of Poisson noise on adiabatic quantum control}
\author{A. Kiely}
\email{anthony.kiely@umail.ucc.ie}
\affiliation{Department of Physics, University College Cork, Cork, Ireland}

\author{J. G. Muga}
\address{Departamento de Qu\'{\i}mica F\'{\i}sica, UPV/EHU, Apdo 644, 48080 Bilbao, Spain}
\address{Department of Physics, Shanghai University, 200444 Shanghai, People's Republic of China}

\author{A. Ruschhaupt}
%\email{aruschhaupt@ucc.ie}
\affiliation{Department of Physics, University College Cork, Cork, Ireland}
%
%
%67.85.-d     Ultracold gases, trapped gases
%42.50.Dv     Quantum state engineering and measurements
%03.65.Aa    Quantum systems with finite Hilbert space
%42.50.-p     Quantum optics
%03.65.-w	Quantum mechanics
%03.65.Xp	Tunneling, traversal time, quantum Zeno dynamics
%03.65.Yz	Decoherence; open systems; quantum statistical methods (see also 03.67.Pp in quantum information; for decoherence in Bose-Einstein condensates, see 03.75.Gg)
%05.40.Ca	Noise
%02.50.Ey	Stochastic processes
\begin{abstract}
We present a detailed derivation of the master equation describing a general time-dependent quantum system with classical Poisson white noise and outline its various properties. We discuss the limiting cases of Poisson white noise and provide approximations for the different noise strength regimes. We show that using the eigenstates of the noise superoperator as a basis can be a useful way of expressing the master equation. Using this we simulate various settings to illustrate different effects of Poisson noise. In particular, we show a dip in the fidelity as a function of noise strength where high fidelity can occur in the strong noise regime for some cases. We also investigate recent claims [Jing et al.,  Phys. Rev. A \textbf{89} 032110 (2014)] that this type of noise may improve rather than destroy adiabaticity.
\end{abstract}
\maketitle

% ---------------------------------------------------------------
% Introduction
% ---------------------------------------------------------------

\section{Introduction}

Understanding the effect of noise is of great interest for creating many of the predicted quantum technologies, e.g., for quantum metrology, quantum cryptography and quantum computation \cite{scheich2016}. Almost all quantum systems suffer from decoherence in one form or another as it is impossible to isolate a system completely. Moreover interactions are needed to prepare, manipulate or read off the state of a system.  Many recent publications have focused on combating different forms of decoherence by designing control schemes which are stable against specific forms of decoherence. Different strategies have been followed to design such schemes, e.g, dynamical decoupling \cite{lidar2012}, composite pulses \cite{torosov2015,ivanov2015,torosov2014}, ``shortcuts to adiabaticity" \cite{sta_review}, and optimal control schemes \cite{glaser2015,frank2015,doria2011}.

There are different possible approaches for modelling this decoherence. One is based on a system-bath theory, where the bath dynamics are traced out under the Born-Markov approximation \cite{breuer2002,carmichael1999}. Another approach is to assume a ``classical noise'', whereby the effect of the bath is described by a stochastic temporal evolution of a closed system. It has been shown for random telegraph noise (also known as a two-state Markov process or dichotomic Markov process) acting on a qubit, that these descriptions lead to equivalent dynamics \cite{saira2007}. Classical noise can of course also occur from classical fluctuations in the experimental system parameters. This noise could also be purposefully used to perform quantum simulations of environmentally induced decoherence \cite{chenu2016}. Hence understanding the effect of classical noise on a quantum system can be quite useful.

In this paper we will consider the effect of classical Poisson white noise (sometimes referred to as white shot noise) \cite{hanggi1980}. It is a sequence of random Markovian strikes with exponential inter-arrival times, i.e., which are Poisson distributed in time. Poisson noise is useful for modeling noise processes which occur as a result of a small number of discrete events, e.g., photons for electromagnetic radiation or electrons for electrical current. For a large number of events, the Poisson distribution tends to a Normal or Gaussian distribution. 

White shot noise has already been widely discussed in the context of classical physics \cite{talkner2007,dykman2010,dykman2013}. It has been applied to a variety of settings, e.g., micro-mechanical resonators \cite{chan2012}, the statistics of current through Josephson junctions \cite{ankerhold_2007,jordan_2009}, modelling random impulsive excitations \cite{khovanov_2011}, and its effects on transport of Brownian particles \cite{hanggi1997,hanggi2013}. It has also been used to model the effect of light intensity fluctuations on photochemical reactions \cite{mccarty1986} and radiation pressure shot noise in optomechanical systems \cite{girvin_2010,regal_2013}. It was first considered in a quantum setting in \cite{niemiec1991} and has since been proposed as a power source for a quantum heat engine \cite{kosloff2012a,kosloff2012b}. It is also a special case of random telegraph noise with vanishing correlation time \cite{broeck1983}. General random telegraph noise has been used to investigate noise effects on tunneling dynamics \cite{pechukas2000},  model the environmental noise of a quantum dot \cite{EntinWohlman2016}, and model decoherence of qubits in general \cite{joynt2008}. A master equation for random telegraph noise has been derived for time independent systems \cite{luczka1991}.

Previous works have mainly focussed on Gaussian noise for stochastic Hamiltonian evolution \cite{budini2001}. Hence, it would be interesting to have a tractable master equation for a more general non-Gaussian noise. Here we will present a general master equation for classical Poisson white noise and show how it simplifies in two-level systems \cite{kosloff2012a,kosloff2012b} and reduces to Gaussian white noise in the appropriate limits \cite{hanggi1980}. 

In a recent paper by Jing et al. \cite{garcia2014} it is claimed that Poisson noise can counterintuitely help improve adiabaticity for increasing noise strength. We will show that what is referred to as strong noise is actually a large noise bias which implies a stronger Hamiltonian. By coherently increasing the energy of the system (for a fixed total time), one will of course improve the adiabaticity. However, we will also show that for a general quantum system with Poisson noise, the system will follow specific eigenstates of the noise superoperator (in the limit of strong noise) in a manner analogous to the adiabatic theorem. This has been previously outlined for the case of Gaussian white noise and has been connected to the effect of repeated measurements or the quantum Zeno effect \cite{zeno,zeno2,zeno3,zeno_adiab,zeno_adiab2}.

The rest of this paper is outlined as follows. In the next section, the master equation for a general time-dependent quantum system with Poisson noise is derived and its general properties discussed including the special case of a two-level system. In Sec. \ref{sect_approx}, we review the adiabatic approximation for density matrices and derive approximations for the cases of weak and strong Poisson noise. In Sec. \ref{sect_examples}, we solve the master equation numerically for several cases, including the setting described in \cite{garcia2014} and Stimulated Raman adiabatic passage (STIRAP) \cite{stirap_review} type schemes in three-level systems. The examples we present will illustrate the different effects of Poisson noise. Finally, in Sec. \ref{Conclusion}, we summarize our work and make some concluding remarks.

\section{Master equation for Poisson noise \label{sect_master}}
We will first derive the master equation for Poisson noise.
Let us consider a Hamiltonian
\begin{eqnarray}
H\left(t\right)=H_{0}\left(t\right)+z\left(t\right)H_{1}\left(t\right),
\end{eqnarray}
where $z\left(t\right)$ is a real function, given by classical Poisson white noise
\begin{eqnarray}
z\left(t\right)=\sum_{i=1}^{N\left(t\right)}\xi_{i}\delta\left(t-t_{i}\right).
\end{eqnarray}
The probability of the number of strikes $N\left(t\right)$ is given by a Poissonian counting process such that the probability
of $n$ strikes after a time $t$ is
\begin{eqnarray}
Q\left(N\left(t\right)=n\right)=\left(\nu t\right)^{n}\frac{e^{-\nu t}}{n!},
\end{eqnarray}
and the random times $t_{i}$ are uniformly distributed on the interval $(0,t)$. The strength of the strikes $\xi_{i}$ are statistically independent of the times and are distributed according to a probability density $P(\xi)$.
The quantity $\nu$ (which corresponds to the quantity $W$ in \cite{garcia2014}) can be thought of as the average frequency of the noise shots. Note that $z(t)$ is dimensionless and the strength of a strike $\xi_{i}$ has dimensions of time. The average and two-time correlation function are given by
\begin{eqnarray}
\avg{z(t)}&=& \nu \avg{\xi}, \\
\avg{z(t)z(s)}-\avg{z(t)}\avg{z(s)}&=& \nu \avg{\xi^2}\delta(t-s).
\end{eqnarray}

For a particular realization of the noise $z\left(t\right)$, the
Liouville-von Neumann equation for the density matrix $\rho_{z}\left(t\right)$ is given by
\begin{eqnarray}
\dot{\rho}_{z}\left(t\right)=-\frac{i}{\hbar}\left[H(t),\rho_{z}\left(t\right)\right]. \label{LvNeq}
\end{eqnarray}
By taking the average over all realizations of $z\left(t\right)$
and defining a new density matrix $\rho\left(t\right)=\left\langle \rho_{z}\left(t\right)\right\rangle _{z}$ this becomes
\begin{eqnarray}
\dot{\rho}\left(t\right)=-\frac{i}{\hbar} \left[H_{0}(t),\rho\left(t\right)\right]
- \frac{i}{\hbar} \left\langle z\left(t\right)\left[H_{1}(t),\rho_{z}\left(t\right)\right]\right\rangle _{z}.
\end{eqnarray}
We now apply the Klyatskin-Tatarsky formula \cite{tatarsky1973,niemiec1991} (one could also consider using the Shapiro-Loginov formula \cite{form_diff}) which has the following form for a Poisson process
\begin{eqnarray}
\lefteqn{\avg{z(t)R[z]}_{z}=} && \nonumber\\
&&\nu \int_{-\infty}^{\infty} d\xi P(\xi) \int_{0}^{\xi}d\eta \avg{\exp\left[\eta \frac{\delta}{\delta z(t)}\right]R[z]}_{z},
\end{eqnarray}
where $R[z]$ is some functional of $z(t)$. In this case $R[z]=\left[H_{1},\rho_{z}\right]$. From Eq.   \eqref{LvNeq}, the functional derivative is
\begin{eqnarray}
\frac{\delta}{\delta z(t)} \rho_{z}(t)=-\frac{i}{\hbar}\left[H_{1}(t),\rho_{z}(t)\right] ,
\end{eqnarray}
and
\begin{eqnarray}
\exp\left[\eta \frac{\delta}{\delta z(t)}\right] \rho_{z}(t)=A_{\eta}\rho_{z}(t) A_{\eta}^{\dagger},
\end{eqnarray}
where $A_{\eta}=e^{-i\eta H_{1}(t)/\hbar}$. From this we arrive at the master equation (where the explicit time dependence has been dropped),
\begin{eqnarray}
\dot{\rho}=-\frac{i}{\hbar}\left[H_{0},\rho\right] +\nu\int_{-\infty}^{\infty}d\xi\,P\left(\xi\right)\left(A_{\xi}\rho A^{\dagger}_{\xi}-\rho\right), 
\label{master1}
\end{eqnarray}
where the following identity has been used
\begin{eqnarray}
\int_{0}^{\xi} d \eta \left[H_{1},A_{\eta}\rho A_{\eta}^{\dagger}\right]=i\hbar\left(A_{\xi}\rho A_{\xi}^{\dagger}-\rho\right).
\end{eqnarray}
Note that Eq. \eqref{master1} is very close to Lindblad form \cite{lindblad_1976}, where the operators $A_{\xi}$ correspond to the Lindblad operators and the sum has been replaced by an integral. By now applying the Hadamard lemma \cite{mandelwolf}, we get the final form of the master equation,
\begin{eqnarray}
\dot{\rho}= \cL_0 (\rho) + \cL_1 (\rho), \label{general_master_eq}
\end{eqnarray}
where
\begin{eqnarray}
\cL_0 (\rho) &=& -\frac{i}{\hbar}\left[H_{0},\rho\right],\label{eq_L0}\\
\cL_1 (\rho) &=& \nu \sum_{s=1}^\infty \frac{1}{s!} \left(-\frac{i}{\hbar}\right)^s \langle \xi^s \rangle \left[ H_1, \rho \right]_s,
\label{eq_L1}
\end{eqnarray}
$\left[ H_1, \rho \right]_s = \left[ H_1, \left[ H_1, \rho \right]\right]_{s-1}$ and $\left[ H_1, \rho \right]_0 = \rho $. Note that $\cL_0$ and $\cL_1$ commute when the two Hamiltonians ($H_{0}$ and $H_{1}$) commute. It is clear from the form of the master equation that it is linear in $\rho$, and by taking the trace of Eq. \eqref{general_master_eq}, we get that $\partial_{t}\tr{\rho}=0$ and hence the trace is preserved.

Gaussian white noise is recovered if one takes the limit $\nu\rightarrow\infty$ such that $\nu\left\langle \xi\right\rangle \rightarrow \widetilde{J}$,
a constant, $\nu\left\langle \xi^{2}\right\rangle \rightarrow2\widetilde{D}$,
a positive constant, and $\nu\left\langle \xi^{s}\right\rangle \rightarrow0\:\forall s>2$ \cite{hanggi1980}. As an explicit example where this happens, let us choose a Laplace distribution
$P\left(\xi\right)=\left(\frac{1}{2A}\right)\exp\left(-\left|\xi\right|/A\right)$
with $A>0$. Since the distribution is symmetric, the odd
moments are 0, i.e., $\left\langle \xi^{2n+1}\right\rangle =0$ for $n \in \mathbb{N}$ and the even ones are given by $\left\langle \xi^{2n}\right\rangle =\left(2n\right)!\,A^{2n}$. From this we can see that $\left\langle \xi\right\rangle =0$ and
that, setting $A=\sqrt{\frac{\widetilde{D}}{\nu}}$, then $\nu\left\langle \xi^{2}\right\rangle =2\widetilde{D}$.
In general we get that $\nu\left\langle \xi^{2n}\right\rangle =\left(2n!\right)\widetilde{D}^{n}\nu^{1-n}$,
hence, $\nu\left\langle \xi^{s}\right\rangle \rightarrow0\:\forall s>2$
as $\nu\rightarrow\infty$.
In this case (and in general taking this limit), the master equation simply reduces to a master equation for Gaussian
white noise,
\begin{eqnarray}
\dot{\rho}=-\frac{i}{\hbar}\left[H_{0}+\widetilde{J} H_{1},\rho\right]-\frac{\widetilde{D}}{\hbar^{2}}\left[H_{1},\left[H_{1},\rho\right]\right],
\label{gaussian_eq}
\end{eqnarray}
which could also be derived directly using Novikov's theorem \cite{novikov1965}.

\subsection{General properties of $\cL_0$ and $\cL_1$ \label{sect_general_prop}}

We will now outline some general properties of $\cL_0$ and $\cL_1$. In the following, the density matrix $\rho(t)$ will be represented as a vector $\Sket{\rho}$ in a larger Hilbert space such that the scalar product is preserved, i.e., for two operators $M_1$ and $M_2$, $\Sbraket{M_1}{M_2} = \tr \left(M_1^{\dagger} M_2\right)$. The equivalence between the two representations will be indicated as $\Sket{\rho} \equiv \rho(t)$.
The superoperators $\cL_0$ and $\cL_1$ can be then seen as linear operators acting on the vector $\Sket{\rho}$.

Let us start by examining $\cL_0$, see Eq. \eqref{eq_L0}. Let $\ket{\phi_n^{(0)}(t)}$ be an instantaneous eigenvector of $H_0$ with eigenvalue $E_n^{(0)}(t)$ and $n \in \bN$ (assuming discrete eigenvalues).
Defining $\Sket{A_{n,m}(t)} \equiv \ketbra{\phi_n^{(0)}(t)}{\phi_m^{(0)}(t)}$, we get
\begin{eqnarray}
\cL_0(t) \Sket{A_{n,m}(t)} = \alpha_{n,m}(t) \Sket{A_{n,m}(t)},
\end{eqnarray}
where $\alpha_{n,m} = -\frac{i}{\hbar} \left(E_n^{(0)}-E_m^{(0)}\right)$ for all $n,m \in \bN$. 
Therefore, $\Sket{A_{n,m}}$ is an eigenvector of the superoperator $\cL_0$ with eigenvalue $\alpha_{n,m}$.
Because the eigenvalues $\alpha_{n,m}$ are purely imaginary, $\cL_0$ is anti-Hermitian, i.e., $\cL_0^{\dagger}=-\cL_0$.

Let us now examine $\cL_1$, see Eq. \eqref{eq_L1}. Let $\ket{\phi_n^{(1)}(t)}$ be an eigenvector of $H_1$ with eigenvalue $E_n^{(1)}(t)$.
Defining $\Sket{B_{n,m}(t)} \equiv \ketbra{\phi_n^{(1)}(t)}{\phi_m^{(1)}(t)}$, we get
\begin{eqnarray}
& &\cL_1  \Sket{B_{n,m}(t)} \nonumber \\
 &=& \nu
\sum_{s=1}^\infty \frac{1}{s!} \left(-\frac{i}{\hbar}\right)^s \langle \xi^s \rangle \left(E_n^{(1)}-E_m^{(1)}\right)^s \Sket{B_{n,m}(t)}\nonumber \\
&=& \beta_{n,m}(t) \Sket{B_{n,m}(t)}.
\end{eqnarray}
Therefore, $\Sket{B_{n,m}}$ is an eigenvector of the superoperator $\cL_1$ with eigenvalue
\begin{eqnarray}
\beta_{n,m} &=& \nu \sum_{s=1}^\infty \frac{1}{s!} \left(-\frac{i}{\hbar}\right)^s \langle \xi^s \rangle \left(E_n^{(1)}-E_m^{(1)}\right)^s\nonumber \\
& = & \nu \left[C_\xi \left(\frac{E_m^{(1)}-E_n^{(1)}}{\hbar}\right)-1\right],\label{L1eigval}
\end{eqnarray}
where  $C_\xi (x) = \langle e^{i\xi x} \rangle$ is the characteristic function of the probability distribution $P(\xi)$.

We now recall some properties of a general characteristic function which are $\fabs{C_\xi (x)} \le 1$, $C_\xi (0) = 1$ and $C_\xi (-x) = C_\xi (x)^*$ for real $x$.
From the last property, it follows that $\beta_{n,m} = \beta_{m,n}^*$.
Moreover, $-2 \nu \le \mbox{Re} (\beta_{n,m}) \le 0$ and $-\nu \le \mbox{Im} (\beta_{n,m}) \le \nu$ for all $n,m$
and $\beta_{n,n} = 0$ for all $n$. For a symmetric  probability distribution, i.e., $P(\xi)=P(-\xi)$, $\cL_1$ is Hermitian and negative. In general, $\cL_{1}$ is always diagonalizable but not necessarily Hermitian.

For numerical treatment it is often useful to represent the master equation in the eigenbasis of $\cL_1$, i.e., $\Sket{\rho}=\sum_{n,m} d_{n,m}\Sket{B_{n,m}}$. Using Eq. \eqref{general_master_eq} we get the following equation for the coefficients of $\Sket{\rho}$ in this basis
\begin{equation} \label{master_eq_basis}
\begin{split}
\dot{d}_{n,m}&-\beta_{n,m}d_{n,m} \\ +&\sum_{i,j}\left[ \delta_{m,j}\left(\braket{\phi_{n}^{(1)}}{\dot{\phi}_{i}^{(1)}}+\frac{i}{\hbar}\bra{\phi_{n}^{(1)}}H_{0}\ket{\phi_{i}^{(1)}}\right)d_{i,j} \right. \\
+& \left. \delta_{n,i}\left(
\braket{\dot{\phi}_{j}^{(1)}}{\phi_{m}^{(1)}}-\frac{i}{\hbar}\bra{\phi_{j}^{(1)}}H_{0}\ket{\phi_{m}^{(1)}}
\right)d_{i,j}\right]=0. 
\end{split}
\end{equation}
In this representation the total contribution from $\cL_1$ arises solely from the eigenvalues $\beta_{n,m}$. The condition for $\rho$ to remain Hermitian is simply $d_{n,m}=d_{m,n}^{*}$ and for it to be pure is $\sum_{n,m}\left|d_{n,m}\right|^{2}=1$. By taking the complex conjugate of Eq.    \eqref{master_eq_basis} we see that $\rho$ will indeed remain Hermitian. So in summary, the master equation is linear and preserves both the trace and Hermiticity. 

\subsection{Special case: Two-level quantum system}
As a special case, consider a two-level quantum system with Hamiltonians given by
\begin{eqnarray}
H_{0}\left(t\right)=\frac{\hbar}{2}\left(\begin{array}{cc}
-\Delta(t) & \Omega_{R}(t)-i\Omega_{I}(t)\\
\Omega_{R}(t)+i\Omega_{I}(t) & \Delta(t)
\end{array}\right), 
\end{eqnarray}
\begin{eqnarray}
H_{1}\left(t\right)=\frac{\hbar}{2}\left(\begin{array}{cc}
-\widetilde{\Delta}(t) & \widetilde{\Omega}_{R}(t)-i\widetilde{\Omega}_{I}(t)\\
\widetilde{\Omega}_{R}(t)+i\widetilde{\Omega}_{I}(t) & \widetilde{\Delta}(t)
\end{array}\right).
\end{eqnarray}
Physically, the Hamiltonian $H_0$ could, for example, correspond to an atom illuminated by a laser which couples only two atomic levels.
In that case, $\Omega_R + i \Omega_I$ would be the Rabi frequency of the coupling and $\Delta$ would be the detuning of the laser.
Possible physical motivations of $H_1$ will be given in the examples in Sect. \ref{sect_examples}.

The eigenvalues of $H_0$ and $H_1$ are $E_\pm^{(0)} =\pm \frac{\hbar}{2} \sqrt{\Omega_R^2+\Omega_I^2+\Delta^2}$ and $E_\pm^{(1)} =\pm \frac{\hbar}{2} \sqrt{\widetilde\Omega_R^2+\widetilde\Omega_I^2+\widetilde\Delta^2}$ respectively. The master equation, Eq. \eqref{master1}, can now be simplified further by applying the Hadamard lemma \cite{mandelwolf} to the integrand of the last term and noticing a recursion relation between nested commutators (see Appendix \ref{app}). We get 
\begin{eqnarray}
\dot{\rho} &=&  -\frac{i}{\hbar}\left[H_{0},\rho\right]
- \frac{D}{\hbar^2}\left[H_{1},\left[H_{1},\rho\right]\right]
-\frac{i}{\hbar} J \left[H_{1},\rho\right]\nonumber\\
&=& -\frac{i}{\hbar}\left[(H_{0} + J H_1),\rho\right]
- \frac{D}{\hbar^2}\left[H_{1},\left[H_{1},\rho\right]\right],
\label{pois_2level_eq} 
\end{eqnarray}
where
\begin{eqnarray}
J &=& \nu \frac{\hbar}{2\sqrt{\chi}}\int_{-\infty}^{\infty}d\xi\,P\left(\xi\right)\sin\left(\frac{2}{\hbar}\xi\sqrt{\chi}\right) \nonumber \\
&=&\nu \sum_{l=0}^{\infty}\frac{\chi^{l}}{\left(2l+1\right)!}2^{2l}\left(-\frac{i}{\hbar}\right)^{2l}\left\langle \xi^{2l+1}\right\rangle ,\\
D &=& \frac{\nu \hbar^2}{2\chi}\int_{-\infty}^{\infty}d\xi\,P\left(\xi\right)\sin^{2}
\left(\frac{1}{\hbar}\xi\sqrt{\chi}\right) \nonumber \\
&=&-\nu \hbar^{2} \sum_{k=1}^{\infty}\frac{\chi^{k-1}}{\left(2k\right)!}2^{2\left(k-1\right)}\left(-\frac{i}{\hbar}\right)^{2k}\left\langle \xi^{2k}\right\rangle,
\end{eqnarray}
and $\chi=\left(E_{\pm}^{(1)}\right)^2$. 
This is the final version of the master equation for Poisson noise in a two-level quantum system. $J$ and $D$ depend on the odd and even moments of $P(\xi)$ respectively. Note that the {\it noise bias} $J$ (which is dimensionless) only modifies the coherent evolution
whereas the {\it noise strength} $D$ (which has dimensions of time) has a decoherent effect. In this case the eigenvalues of the superoperator $\cL_1$ (see Eq. \eqref{L1eigval}) are given by
\begin{eqnarray}
\beta_{n,m} = -\frac{i}{\hbar} J (E_n^{(1)}-E_m^{(1)}) - \frac{D}{\hbar^2} (E_n^{(1)}-E_m^{(1)})^2,
\end{eqnarray}
where $n = \pm$ and $ m = \pm$.

The master equation for a two-level system with Poisson noise has the same form as the case of Gaussian white noise (see Eq. \eqref{gaussian_eq}) apart from different expressions for the constant coefficients $J$ and $D$.
In the limit in which Poisson noise converges to Gaussian noise, then $J \to \widetilde J$ and $D \to \widetilde D$.

% ---------------------------------------------------------------------------------------
% ---------------------------------------------------------------------------------------
% ---------------------------------------------------------------------------------------

\section{Approximations for weak and strong Poisson noise \label{sect_approx}}

In this section we consider the different regimes of adiabaticity with no noise, weak noise and strong noise.

\subsection{Adiabatic approximation without noise}
We will first review the adiabatic approximation without noise. The master
equation is then
\begin{eqnarray}
\frac{d}{dt} \Sket{\rho(t)} = \cL_0(t) \Sket{\rho(t)}.
\label{master0}
\end{eqnarray}
We are interested in the dynamics for a slowly varying $\cL_{0}$, i.e., for large total time $T$.
In the usual adiabatic approximation for the Schr\"odinger equation with an initial state $\ket{\psi(0)}=\sum_{n} a_{n} \ket{\phi_{n}^{(0)}(0)}$, the state evolves as
\begin{equation}
\begin{split}
\ket{\psi(T)} \approx& \ket{\psi_{ad} (T)} = \sum_{n} a_{n} \exp \left[-\frac{i}{\hbar}\int_{0}^{T}ds \,E_{n}^{(0)}(s) \right. \\  
-& \left. \int_{0}^{T}ds \braket{\phi_{n}^{(0)}(s)}{ \dot{\phi}_{n}^{(0)}(s)}\right]\ket{\phi_{n}^{(0)}(T)}
\end{split}
\label{puread}
\end{equation}
for large $T$. To simplify the notation, we will now assume that the time-dependent phase of $\ket{\phi_{n}^{(0)}(t)}$ has been
chosen such that $\braket{\phi_{n}^{(0)}(t)}{ \dot{\phi}_{n}^{(0)}(t)} = 0$ for all $n$ and $t$, i.e., the parallel transport condition. This condition can always be fulfilled.
While it is always true that $\Sbraket{A_{n,n}(t)}{ \dot{A}_{n,n}(t)} = 0$, with this
assumption about $\ket{\phi_{n}^{(0)}(t)}$, it also follows $\Sbraket{A_{n,m}(t)}{ \dot{A}_{n,m}(t)} = 0$ for all $n,m$.

Motivated by Eq. \eqref{puread}, we now use the ansatz
\begin{equation}
\Sket{\rho (t)} = \sum_{n,m} b_{n,m} (t) \exp \left[\Lambda_{n,m} (t)\right] \Sket{A_{n,m}(t)},
\end{equation}
for the density matrix, where $b_{n,m}(t)$ are time-dependent coefficients and
\begin{eqnarray}
\Lambda_{n,m} (t) = \int_0^t ds \, \alpha_{n,m} (s).
\end{eqnarray}
Inserting this into Eq. \eqref{master0}, it follows that
\begin{eqnarray}
\lefteqn{\dot{b}_{n,m}(t)=}&&  \nonumber \\ 
&&-\sum_{\substack{l,k \\ (l,k) \neq (n,m)}} \exp \left [\Lambda_{l,k}(s)-\Lambda_{n,m}(s) \right] \Sbraket{A_{n,m}}{\dot{A}_{l,k}} b_{l,k}(t). \nonumber \\
\end{eqnarray}
By assuming a large value of $T$ and following similar steps as in the derivation of the adiabatic approximation for
pure states, we get that $b_{n,m}(T) \approx b_{n,m} (0) = \Sbraket{A_{n,m}(0)}{\rho(0)}$.
Therefore the adiabatic approximation is
\begin{equation}
\Sket{\rho (T)} \approx  \sum_{n,m} b_{n,m} (0) \exp \left[ \Lambda_{n,m} (T)\right] \Sket{A_{n,m}(T)} .
\end{equation}

Let us consider that the system starts in a pure state $\ket{\psi(0)} = \sum_n a_n \ket{\phi_{n}^{(0)}(0)}$ (where $\sum_{n} \left| a_{n}\right|^{2}=1$). It follows that
$\Sket{\rho(0)} \equiv \ketbra{\psi(0)}{\psi(0)}$ and so
$b_{n,m}(0) = a_n a_m^*$. Then,
\begin{eqnarray}
\Sket{\rho (T)} &\approx&  \sum_{n,m} a_n a_m^* \exp \left[ \Lambda_{n,m}(T) \right]
\Sket{A_{n,m}(T)} \nonumber \\
&  \equiv & \ketbra{\psi_{ad} (T)}{\psi_{ad} (T)},
\end{eqnarray}
where $\ket{\psi_{ad} (T)}$ is given in Eq. \eqref{puread}.
If the system starts in an energy eigenstate of $H_0$, we get that
$\Sket{\rho(0)} = \Sket{A_{N,N}(0)} \equiv \ketbra{\phi_{N}^{(0)}(0)}{\phi_{N}^{(0)}(0)}$ for a fixed $N$.
It follows $b_{n,m} (0) = \delta_{n,N} \delta_{m,N}$.
%and also $\Sbraket{A_{N,N}(s)}{\partial_s A_{N,N}(s)} = 0$.
Therefore the adiabatic approximation becomes
\begin{eqnarray}
\Sket{\rho (T)} \approx \Sket{A_{N,N}(T)}
\end{eqnarray}
since $\alpha_{N,N} (t) = 0$.

\subsection{Approximation for weak noise in an adiabatic process}

In this section, we will consider the effect of weak Poisson noise on an adiabatic process. We start with the general master equation for Poisson noise
\begin{eqnarray}
\frac{d}{dt} \Sket{\rho(t)} = \left[\cL_0(t) + \kappa \cL_1(t)\right] \Sket{\rho(t)},
\end{eqnarray}
where we have included a dimensionless coefficient $\kappa$ which is an auxiliary variable used to perform a series expansion. It corresponds to the strength of the noise superoperator $\cL_1$ and will be assumed to be a small quantity in this section.

We assume that the system starts at $t=0$ in a pure state $\Sket{\rho(0)} \equiv \ketbra{\psi(0)}{\psi(0)}$,
where  $\ket{\psi(0)} = \sum_n a_n \ket{\phi_n^{(0)}(0)}$.
It should end at $t=T$ in the state $\Sket{\rho_{ad}} \equiv \ketbra{\psi_{ad}(T)}{\psi_{ad}(T)}$.
We define a fidelity $F$, such that $F^2= \Sbraket{\rho_{ad}}{\rho(T)}= \bra{\psi_{ad}(T)}\rho(T)\ket{\psi_{ad}(T)}$.
We can expand this in terms of the small quantity $\kappa$ to get the approximation
\begin{eqnarray}
F(\kappa) \approx F(0) + \kappa F'(0),
\label{noiseapprox}
\end{eqnarray}
where the noise sensitivity is 
\begin{eqnarray}
F'(0) = \frac{1}{2 F(0)} \int_0^T dt \Sbra{\widetilde{\rho} (t)} \cL_1(t) \Sket{\rho_{0}(t)}.
\end{eqnarray}
We have defined $\Sket{\widetilde \rho (t)} = \cU_0 (t,T) \Sket{\rho_{ad}}$ and $\Sket{\rho_{0}(t)}=\cU_0 (t,0) \Sket{\rho(0)}$, where $\cU_0 (t_{2},t_{1})=\mathcal{T} \exp\left[\int_{t_{1}}^{t_{2}}ds \cL_{0}(s)\right]$ is the noise-less time-evolution operator and $\mathcal{T}$ is the time ordering operator. Note that we do not assume perfect adiabatic transfer in the unperturbed case.

If the system starts at $t=0$ in an energy eigenstate of $H_0$, i.e., $\Sket{\rho(0} = \Sket{A_{NN}(0)} \equiv \ketbra{\phi_N^{(0)}(0)}{\phi_N^{(0)}(0)}$, the target state is
$\Sket{\rho_{ad}} = \Sket{A_{NN}(T)} \equiv \ketbra{\phi_N^{(0)}(T)}{\phi_N^{(0)}(T)}$.
In this case, the noise sensitivity is
\begin{eqnarray}
F'(0) = \frac{1}{2 F(0)} \int_0^T dt \Sbra{\widetilde A_{NN} (t)} \cL_1(t) \Sket{\rho_{0}(t)},
\label{sensitivity}
\end{eqnarray}
where $\Sket{\widetilde A_{NN} (t)} = \cU_0 (t,T) \Sket{A_{NN} (T)}$.
In the following examples, the noise sensitivity $F'(0)$ is negative. This shows that in these cases a small amount of noise will not improve the fidelity, contrary to the claim in \cite{garcia2014}.

\subsection{Strong noise limit \label{strong_sec}}

In this section, we will consider the case of strong noise, i.e., where $\cL_{1}$ is dominant. Note that this is not the same as the Gaussian noise limit. The master equation is once again given by
\begin{eqnarray}
\frac{d}{dt} \Sket{\rho(t)}=\left[\cL_{0}(t)+\kappa \cL_{1}(t)\right]\Sket{\rho(t)},
\label{strongmaster}
\end{eqnarray}
where $\kappa$ is again an auxiliary variable (which corresponds to the strength of the superoperator $\cL_1$) used for the purposes of approximation. In this case it will be assumed to be large.
A discussion of the adiabatic condition for non-unitary evolution can be found in \cite{ibanez2014}.
However the setting in Eq. \eqref{strongmaster} differs from this in the sense that only part of the right-hand side is dominant. Note that $\cL_0(t)$ and $\cL_1(t)$ can always be diagonalized (see Sec. \ref{sect_general_prop}). The case of an adiabatic approximation where the superoperator can only be tranformed in a Jordan canonical form can be found in \cite{lidar}.

Recall that the instantaneous eigenvectors of $\cL_1$ are $\Sket{B_{n,m}} \equiv \ketbra{\phi_n^{(1)}}{\phi_m^{(1)}}$ with corresponding eigenvalues $\beta_{n,m}$ (see Sec. \ref{sect_general_prop}).
To simplify the notation, we will assume that $\braket{\phi_{n}^{(1)}(t)}{ \dot{\phi}_{n}^{(1)}(t)} = 0$ for all $n$ and $t$. It then follows that $\Sbraket{B_{n,m}(t)}{ \dot{B}_{n,m}(t)} = 0$ for all $n,m$. Moreover, we assume a symmetric probability distribution $P(\xi)$ which results in real negative eigenvalues $\beta_{n,m}$ and $\cL_1$ Hermitian. While it is always the case that $\beta_{n,n} = 0$, we also assume that $\beta_{n,m} = 0$ if and only if $n=m$. This is fulfilled if the eigenvalues of $H_1$ are non-degenerate and $C_\xi (x) = 1$ if and only if $x=0$.

If the initial state is expressed as $\Sket{\rho(0)} = \sum_{n,m} c_{n,m} (0)\Sket{B_{n,m}(0)}$ (where $c_{n,m}(0)=\Sbraket{B_{n,m}(0)}{\rho(0)}$), then motivated by the usual adiabatic theorem in quantum mechanics and by \cite{lidar} we use the general ansatz
\begin{eqnarray}
\Sket{\rho(t)} = \sum_{n,m} c_{n,m} (t) \exp \left[\tilde \Lambda_{n,m} (t) \right] \Sket{B_{n,m}(t)}, \label{ansatzS}
\end{eqnarray}
where
\begin{eqnarray}
\lefteqn{\tilde\Lambda_{n,m} (t) =} && \\
&& \int_0^t ds\, \left[\kappa \beta_{n,m}(s) + \Sbraket{B_{n,m}(s)}{\cL_0|B_{n,m}(s)}\right]. \nonumber
\end{eqnarray}
If we now insert this into Eq. \eqref{strongmaster}, we get that
\begin{eqnarray}
\lefteqn{\dot{c}_{n,m}(t)=}&&\nonumber \\ &&\sum_{\substack{l,k \\ (l,k)\neq(n,m)}} \exp \left [ \tilde{\Lambda}_{l,k}(t)-\tilde{\Lambda}_{n,m}(t) \right]  M_{n,m,l,k}(t)c_{l,k}(t), \label{ceqstrong} \nonumber \\
\end{eqnarray}
where
\begin{eqnarray}
M_{n,m,l,k}(t)=\Sbraket{B_{n,m}}{\cL_0|B_{l,k}}-\Sbraket{B_{n,m}}{\dot{B}_{l,k}}.
\end{eqnarray}
For large noise $\kappa$ (see Appendix \ref{app2} for details)
\begin{eqnarray}
\Sket{\rho(t)} \approx \sum_{n,m} c_{n,m} (0) \exp \left[\tilde\Lambda_{n,m}(t) \right] \Sket{B_{n,m}(t)}. 
\label{midstep}
\end{eqnarray}
Note that $\tilde\Lambda_{n,n}(t)=0$. If $n \neq m$, $\exp\left[\tilde\Lambda_{n,m}(t)\right] \to 0$ in the limit of $\kappa \rightarrow \infty$. Hence the final result is
\begin{eqnarray}
\Sket{\rho(t)} &\approx & \sum_{n} c_{n,n} (0)\Sket{B_{n,n}(t)} \nonumber \\
                 &=& \Sket{\rho_{\infty}(t)} .\label{limitcase}
\end{eqnarray}
We define the strong noise limit fidelity $F_{\infty}$ as $F_{\infty}^2= \Sbraket{\rho_{\infty}}{\rho}$. The only remaining elements are those which are not affected by $\cL_1$, i.e., $\cL_1 \Sket{B_{n,n}}=0$. These are the diagonal elements of the density matrix in the eigenbasis of $H_{1}$. For example, if $H_{1}=H_{0}$, the noise term simply projects on the eigenstates of $H_{0}$. Hence, if the state starts in an eigenstate of $H_{0}$, it will remain in that eigenstate in the strong noise regime. However, a superposition of eigenstates will not survive, as the noise term clearly kills any coherence terms (or off-diagonal elements of the density matrix).
This is different from the adiabatic approximation applied to $\cL_0$ for large time in a previous subsection. 

The purity of the general ansatz (Eq. \eqref{ansatzS}) becomes
\begin{eqnarray}
\Sbraket{\rho(t)}{\rho(t)} \rightarrow \sum_{n} c_{n,n}(0)^{2} 
\end{eqnarray}
in the limit $\kappa \rightarrow \infty$. The system will remain in a pure state in the strong noise limit if the density matrix is diagonal in the $H_{1}$ eigenbasis at $t=0$.

% ---------------------------------------------------------------------------------------
% ---------------------------------------------------------------------------------------
% ---------------------------------------------------------------------------------------

\section{Poisson noise effect on adiabaticity \label{sect_examples}}

In this section, we will present different types of effects of Poisson noise on adiabaticity using several illustrating examples.

\subsection{Phase-changing scheme in a two-level system}

We start by examining the setting which is also considered in \cite{garcia2014},
i.e., a two-level quantum system with Poisson white noise. While the Poisson noise used in \cite{garcia2014} is always Gaussian, we will continue to use the notation for Poisson white noise since obtaining the results for Gaussian white noise only requires a relabelling $J \to \widetilde J$ and $D \to \widetilde D$. The noise Hamiltonian is $H_1=H_{0}$ such that the master equation is
\begin{eqnarray}
\dot{\rho}=-\frac{i}{\hbar}\left[(1+J) H_{0},\rho\right]-\frac{D}{\hbar^{2}}\left[H_{0},\left[H_{0},\rho\right]\right].
\label{mastersimple}
\end{eqnarray}
Instead of averaging over different realizations of the noise as is done in \cite{garcia2014}, we will directly solve this master equation numerically. This avoids any convergence issues  that could arise when numerically averaging over multiple realizations. We use the following scheme from \cite{garcia2014}:
\begin{eqnarray}
\Omega_R(t) = 2 \Omega_{0} \fcos{\Omega t}, \,
\Omega_I(t) = 2 \Omega_{0} \fsin{\Omega t}, \,
\Delta = - \Omega_{0}. \nonumber \\
\end{eqnarray}
This scheme only changes the relative phase of the state and not the populations. The goal is to follow adiabatically the eigenstate $\ket{\phi_{+}^{(0)}}$ of $H_{0}$.

%%%%%%%%%%%%%%%%%%%%%%%%%%%%%%%%%%%%%%%%%%
%Figure
%%%%%%%%%%%%%%%%%%%%%%%%%%%%%%%%%%%%%%%%%%
\begin{figure}[t]
\begin{center}
(a) \includegraphics[angle=0,width=0.7\linewidth]{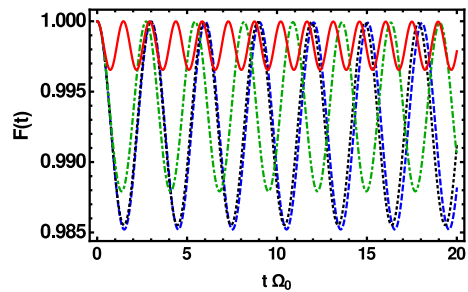} \\[0.5cm]
(b) \includegraphics[angle=0,width=0.7\linewidth]{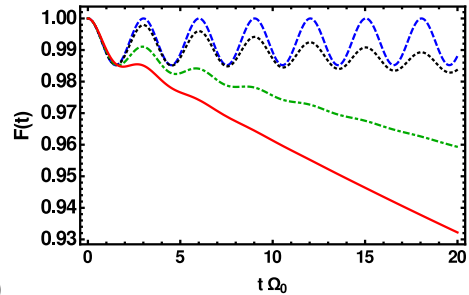}

\end{center}
\caption{Phase changing scheme. Fidelity $F(t)$ versus time $t$.
(a) $J=0$ (blue, dashed line), $J=0.01$ (black, dotted line), $J=0.1$ (green, dot-dashed line), $J=1$ (red, solid line), $D \Omega_{0}=10^{-4}$ in all cases.
(b) $J=0$; $D=0$ (blue, dashed line), $D \Omega_{0}=0.01$ (black, dotted line), $D \Omega_{0}=0.05$ (green, dot-dashed line), $D\Omega_{0}=0.1$ (red, solid line). $\Omega=0.4 \, \Omega_{0}$ and $T \Omega_{0}=20$.}  
\label{fig_phase}
\end{figure}
%%%%%%%%%%%%%%%%%%%%%%%%%%%%%%%%%%%%%%%%%%
%Figure
%%%%%%%%%%%%%%%%%%%%%%%%%%%%%%%%%%%%%%%%%%

We now simulate the master equation (Eq. \eqref{mastersimple}) and plot the fidelity $F(t)=\sqrt{\bra{\phi_{+}^{(0)}(t)} \rho(t) \ket{\phi_{+}^{(0)}(t)}}$
versus time. In Fig. \ref{fig_phase}(a), the fidelity is plotted for different values of the noise bias $J$ with a very small noise strength $D \Omega_{0}=10^{-4}$.
The corresponding plot is qualitatively similar to Fig. 1 in \cite{garcia2014} as we have used similar parameter values.
The fidelity increases with increasing noise bias $J$ for a fixed, small noise strength $D$ (this is also discussed in detail in \cite{garcia2014}).
This can be easily understood from the master equation (Eq. \eqref{mastersimple}); increasing $J$ (with a fixed and almost negligible $D$)
has just the same effect as increasing the strength of the Hamiltonian $H_0$ which clearly results in a better adiabatic behavior.

The outcome is completely different if we fix $J=0$ and increase $D$. This can be seen in Fig. \ref{fig_phase}(b).
Good adiabaticity, i.e., high fidelity, is only found for small $D$. In general the fidelity is decreasing with increasing $D$. This agrees with the natural intuition that noise typically destroys adiabaticity.
In the following, the effect of the noise strength $D$ on adiabatic schemes is investigated further. From this point on, $J=0$ always since it only changes the coherent evolution.

\subsection{Population transfer in a two-level system}

In this section, we continue to consider a two-level system but now for a population transfer scheme. We assume the following
Rapid Adiabatic Passage(RAP) protocol \cite{adiabP1,adiabP2,adiabP3}
\begin{eqnarray}
\Omega_{R}(t)&=& \Omega_{0} \sin\left(\frac{\pi t}{T}\right), \nonumber \\
\Omega_{I}(t)&=& 0, \nonumber \\
\Delta(t)&=& -\delta_{0} \cos\left(\frac{\pi t}{T}\right),
\end{eqnarray}
which produces a population inversion in the bare basis. The system starts in an instantaneous energy eigenstate $\ket{\phi_{+}^{(0)}(t)}$ of $H_{0}(t)$. We use the same definition of fidelity as in the previous subsection.
We now simulate Eq. \eqref{pois_2level_eq}.

%%%%%%%%%%%%%%%%%%%%%%%%%%%%%%%%%%%%%%%%%%
%Figure
%%%%%%%%%%%%%%%%%%%%%%%%%%%%%%%%%%%%%%%%%%
\begin{figure}[t]
\begin{center}
(a) \includegraphics[angle=0,width=0.7\linewidth]{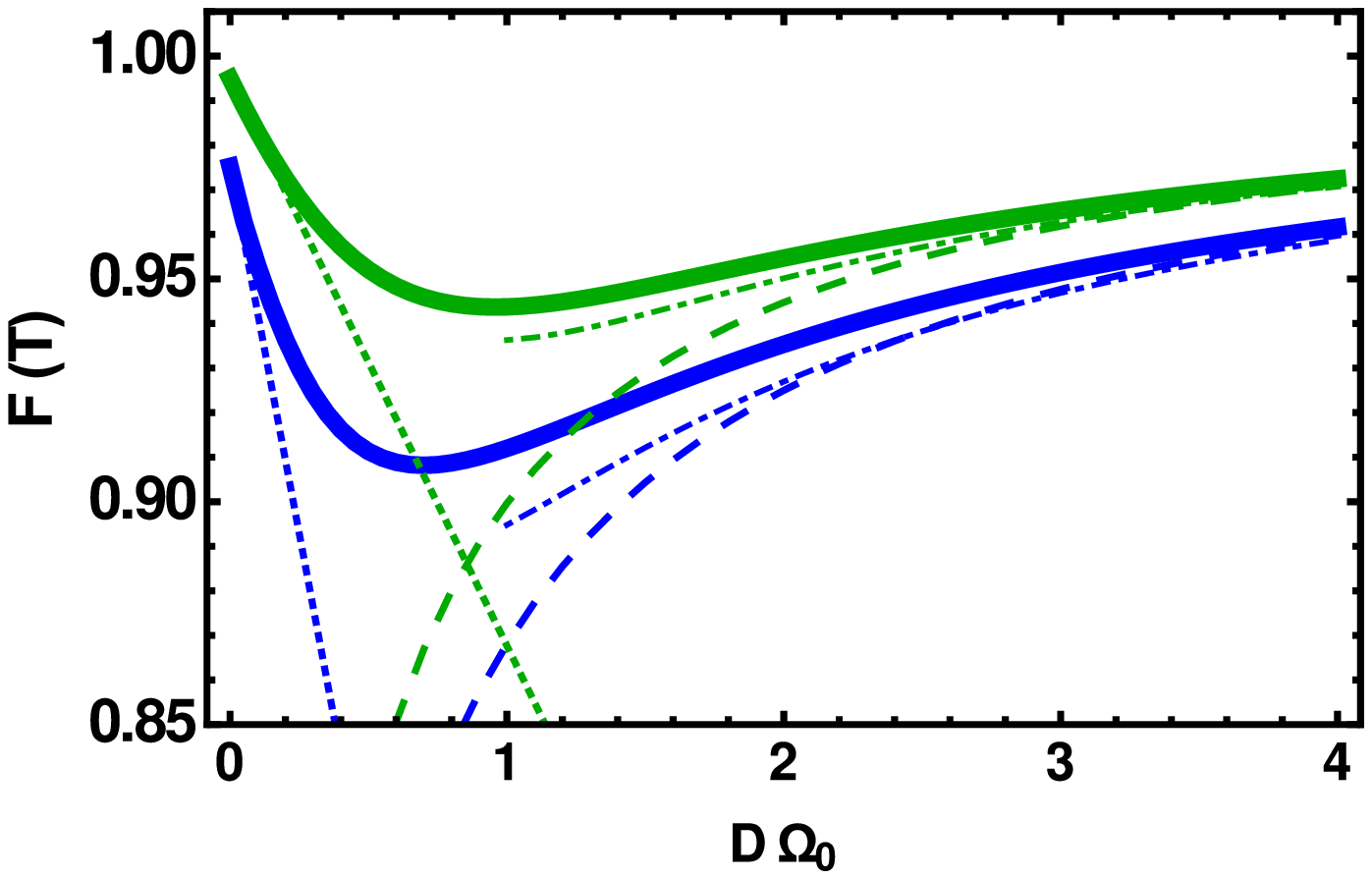} \\[0.5cm]
(b) \includegraphics[angle=0,width=0.9\linewidth]{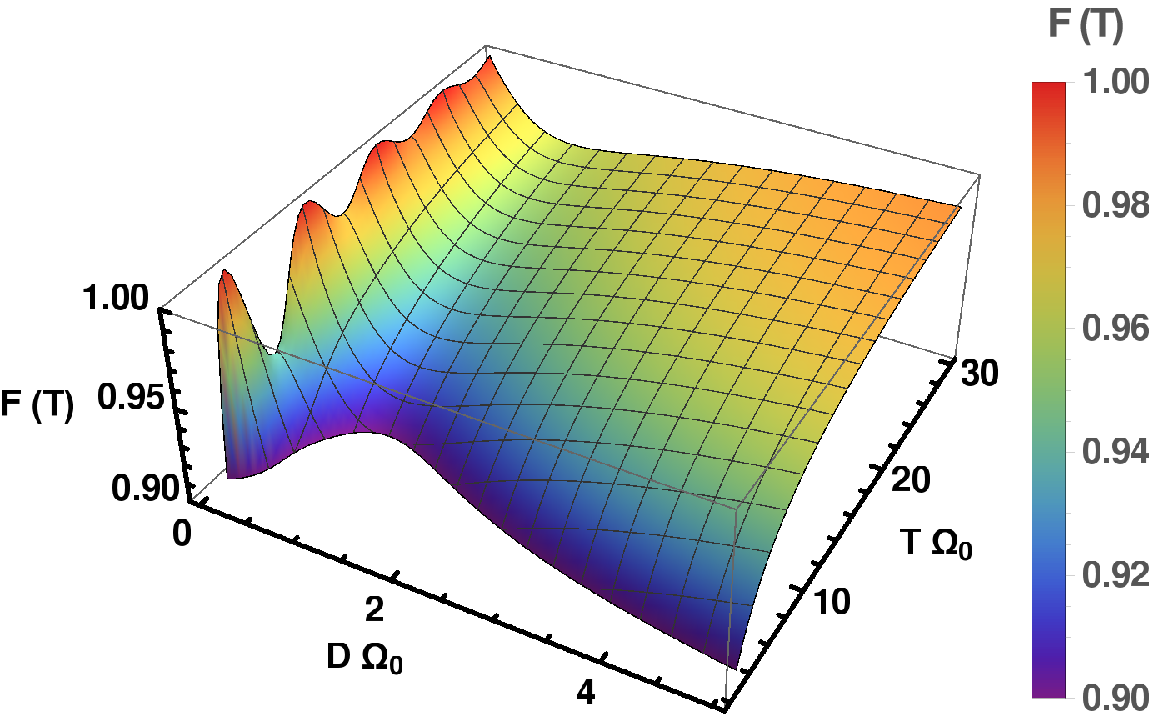}
\end{center}
\caption{RAP scheme in a two-level system with $H_0 = H_1$. (a) Fidelity $F(T)$ against noise strength $D$,
$\delta_0 = 3.5 \, \Omega_{0}$(blue) and
$\delta_0 =1 \, \Omega_{0}$(green). Numerically exact solution (solid lines), small noise approximation Eq. \eqref{noiseapprox} (dotted lines), Naive strong noise approximation (dashed lines) and strong noise approximation Eq. \eqref{secondorderapprox} (dot-dashed line);
$T \Omega_{0}=20 $.
(b) Fidelity $F(T)$ against both noise strength $D$ and total time $T$;
$\delta_{0}=1 \, \Omega_{0}$.
 }
\label{fig_rap}
\end{figure}
%%%%%%%%%%%%%%%%%%%%%%%%%%%%%%%%%%%%%%%%%%
%Figure
%%%%%%%%%%%%%%%%%%%%%%%%%%%%%%%%%%%%%%%%%%

%%%%%%%%%%%%%%%%%%%%%%%%%%%%%%%%%%%%%%%%%%
%Figure
%%%%%%%%%%%%%%%%%%%%%%%%%%%%%%%%%%%%%%%%%%
\begin{figure}[t]
\begin{center}
(a) \includegraphics[angle=0,width=0.7\linewidth]{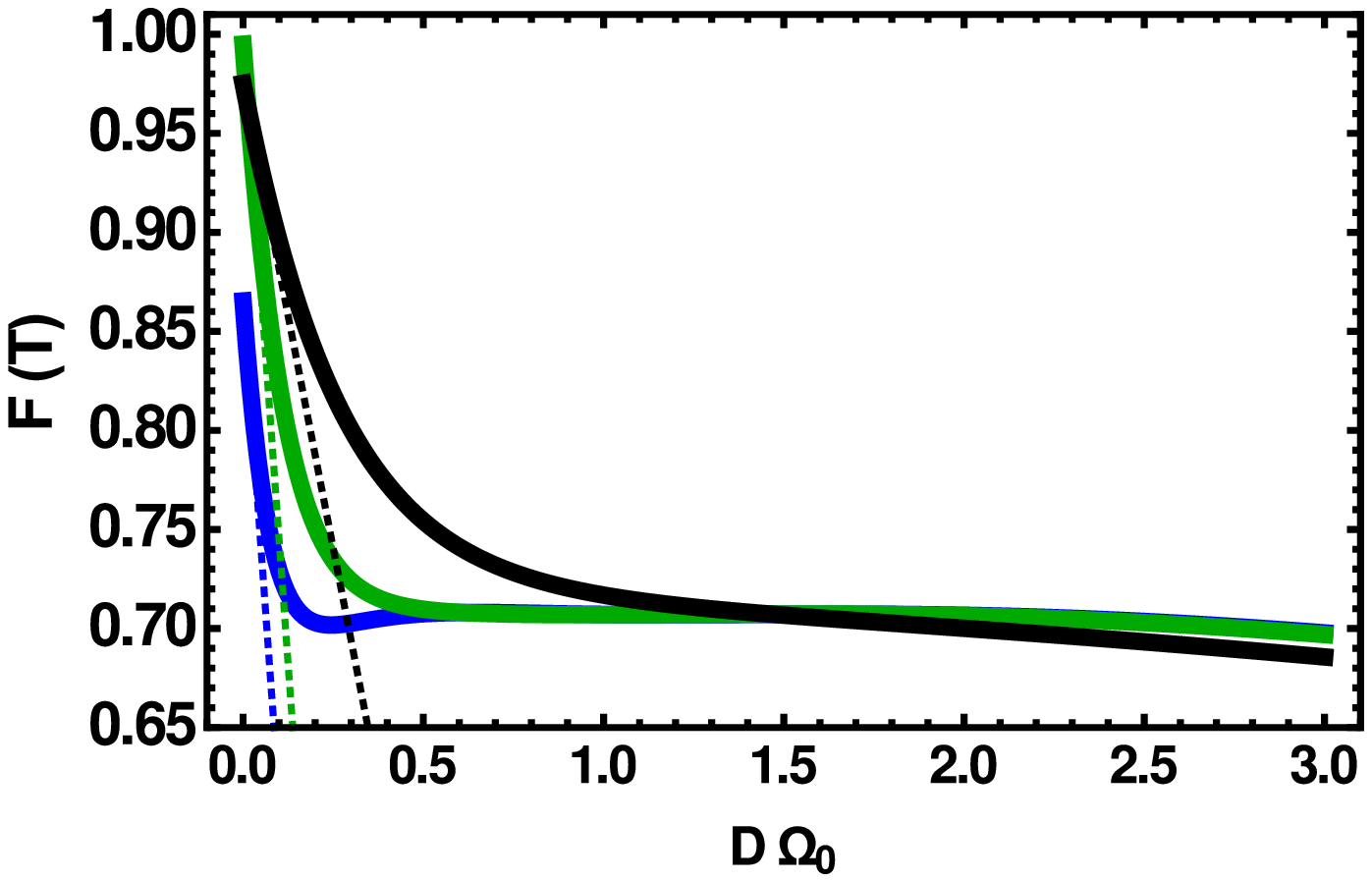} \\[0.5cm] %\hspace*{0.5cm}%
(b) \includegraphics[angle=0,width=0.7\linewidth]{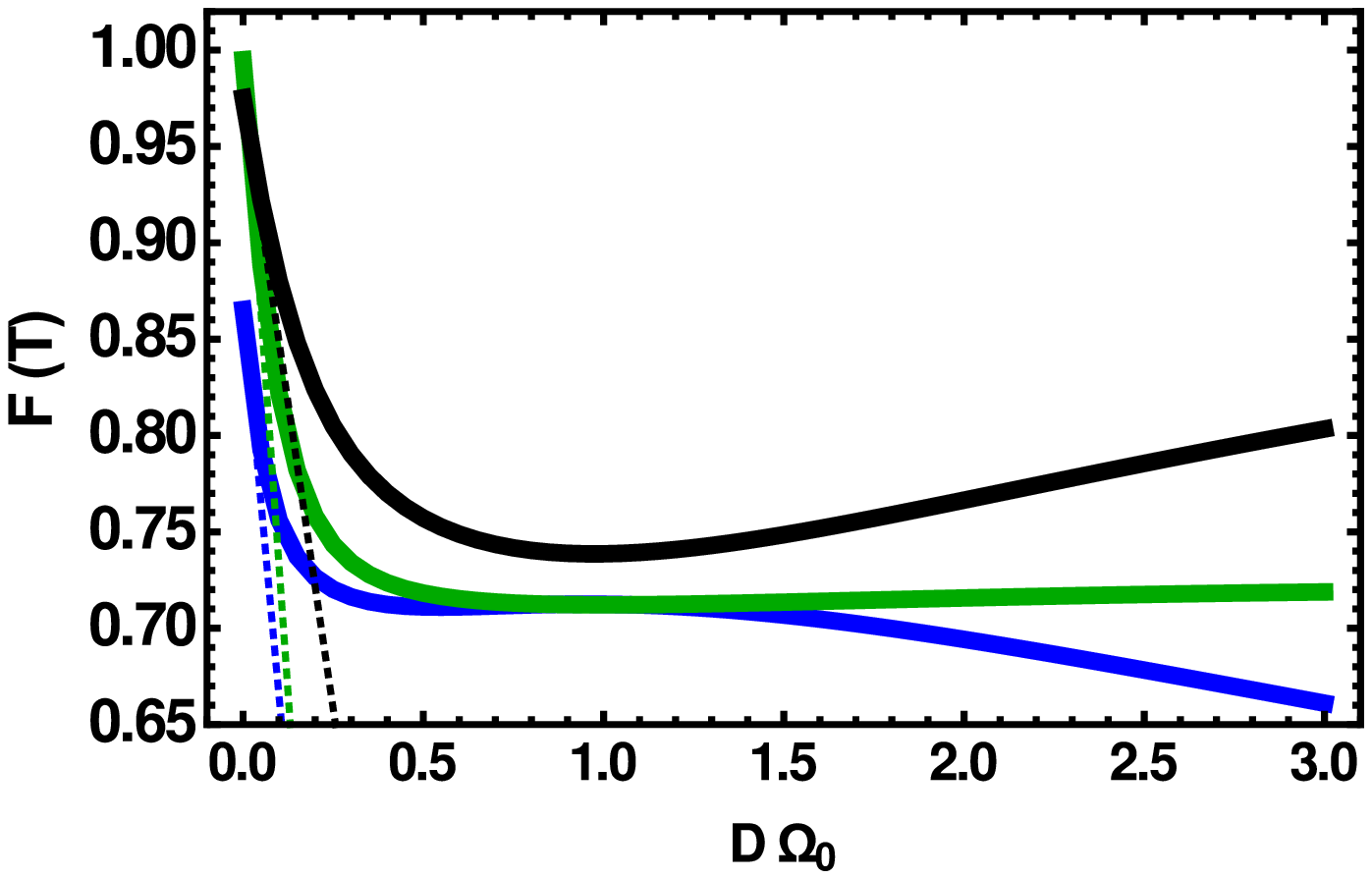}% \\[0.5cm]
\end{center}
\caption{RAP scheme in a two-level system with $H_0 \neq H_1$. $ \delta_0 = 3.5 \, \Omega_0$(black), $ \delta_0 = 1 \,\Omega_0$(green), $ \delta_0 = 0.5 \, \Omega_0$(blue).
Numerically exact solution (solid lines) and
small noise approximation Eq. \eqref{noiseapprox} (dotted lines). (a) Fidelity $F(T)$ against noise strength $D$ for frequency error.
(b) Fidelity $F(T)$ against noise strength $D$ for both timing and frequency error with $c=1$. }
\label{fig_rapx}
\end{figure}
%%%%%%%%%%%%%%%%%%%%%%%%%%%%%%%%%%%%%%%%%%
%Figure
%%%%%%%%%%%%%%%%%%%%%%%%%%%%%%%%%%%%%%%%%%

First consider the case $H_1 = H_0$. Physically, this could originate from
Poisson noise in the total Hamiltonian, or from Poisson noise in the timing
of the process.

In Fig. \ref{fig_rap}(a), the fidelity is decreasing for small noise,
i.e., the noise sensitivity (see Eq. \eqref{sensitivity}) is negative. This shows that
a small amount of noise will not improve the fidelity as one would expect.
The fidelity is decreasing with increasing noise strength $D$. 

However, at some point the fidelity begins to increase again due to the effect of strong noise.
Even though the noise bias $J$ is zero, there is a convergence to the strong noise limit fidelity $F_{\infty}=1$ in this case. The strong noise approximation Eq.   \eqref{secondorderapprox} (which is only plotted in the strong noise regime $D \geq 1$) is compared to the naive strong noise solution (which is the solution of the equation $\dot{\rho}=\cL_{1}\left(\rho\right)$). While only heuristic, this naive approach works well in the limit of strong noise. However, it is clearly not as accurate as the approach presented in Appendix \ref{app2}.

In Fig. \ref{fig_rap}(b), the same fidelity plotted against both noise strength $D$ and total time $T$ is shown.
For $D=0$, the fidelity oscillates slowly towards $1$ for increasing $T$, i.e., the adiabatic limit.
In general the fidelity increases both for increasing $T$ (adiabatic limit) and increasing $D$ (strong noise limit).
The same dip in Fig. \ref{fig_rap}(a) is present here also. 

We also consider examples where $H_1 \neq H_0$. Firstly we consider an absolute error in the detuning which could be due to an error in the laser frequency. In this case
\begin{eqnarray}
\widetilde{\Omega}_R(t) = 0, \,
\widetilde{\Omega}_I(t) = 0, \,
\widetilde{\Delta}(t) = \Omega_{0}. 
\end{eqnarray}

This is shown in Fig. \ref{fig_rapx}(a). For all values of $\delta_{0}$ the fidelity decreases for increasing noise strength. In particular the value of the fidelity in the strong noise limit is $F_{\infty}=0$ for all cases. However there are some cases whereby the fidelity can increase again for large noise strengths even though $H_1 \neq H_0$.

One possible example of this is a case where there is both noise in the detuning $z(t)$ and noise in the timing of the process $\tilde{z}(t)$. In this case we assume that the different noises are proportional $z(t)=c \tilde{z}(t)$ and ignore higher order terms to get a noise Hamiltonian
\begin{eqnarray}
H_{1}\left(t\right)=\frac{\hbar}{2}\left(\begin{array}{cc}
-(\Delta(t)+c \widetilde{\Delta}) & \Omega_{R}(t)-i\Omega_{I}(t)\\
\Omega_{R}(t)+i\Omega_{I}(t) & \Delta(t)+c \widetilde{\Delta}
\end{array}\right).
\end{eqnarray}

 In Fig. \ref{fig_rapx}(b), we can see the fidelity is plotted against noise strength $D$ for $c=1$. The examples shown represent $c \widetilde{\Delta}<\delta_{0}$, $c\widetilde{\Delta}=\delta_{0}$ and $c\widetilde{\Delta}>\delta_{0}$. The limiting solution for $c\widetilde{\Delta}>\delta_{0}$ is $F_{\infty}=1$ as $H_{0}$ and $H_{1}$ have the same eigenvectors at initial and final times, i.e., $t=0$ and $t=T$. The limiting solution for $c\widetilde{\Delta}<\delta_{0}$ is $F_{\infty}=0$ since $H_{0}$ and $H_{1}$ have eigenvectors which are exactly opposite at the initial time but the same at the final time. If $c\widetilde{\Delta}=\delta_{0}$, there is a degeneracy in $H_{1}$ at $t=0$. This leads to a maximally mixed state in the strong noise limit with $F_{\infty}=1/\sqrt{2}$. 
 
While this example is perhaps not the most realistic (since it is assumed that both noise terms are proportional and higher terms can be neglected), it provides a nice example of the different possible effects noise may have on the fidelity. In particular it is possible to achieve high fidelity for strong noise even when $H_{1} \neq H_{0}$.

In the two-level model, the previous results can be also applied if the Poisson noise becomes Gaussian noise because the change from Poisson noise to Gaussian noise just corresponds to a reinterpretation  $J \to \widetilde J$ and $ D \to \widetilde D$.
A third example, using a more complex quantum system will be considered in the next subsection. The master equation for Poisson white noise
will no longer be of the same form as that for Gaussian white noise.

\subsection{STIRAP process in a three-level system}

%%%%%%%%%%%%%%%%%%%%%%%%%%%%%%%%%%%%%%%%%%
%Figure
%%%%%%%%%%%%%%%%%%%%%%%%%%%%%%%%%%%%%%%%%%
\begin{figure}[t]
\begin{center}
\includegraphics[angle=0,width=0.7\linewidth]{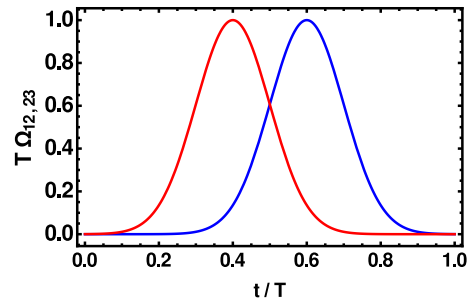} 

\end{center}
\caption{STIRAP pulse sequence with $T\Omega_{0}=1$ and $\tau\Omega_{0}=0.1$; $\Omega_{12}$(blue) and $\Omega_{23}$(red).}  
\label{fig_stirap_pulse}
\end{figure}
%%%%%%%%%%%%%%%%%%%%%%%%%%%%%%%%%%%%%%%%%%
%Figure
%%%%%%%%%%%%%%%%%%%%%%%%%%%%%%%%%%%%%%%%%%

Consider now a three-level quantum system and a STIRAP scheme for population transfer. In this setting the master equation for Poisson noise does not simplify to a form similar to the Gaussian noise master equation. The Hamiltonian is now
\begin{eqnarray}
H_0(t) &=& \frac{\hbar}{2} \left(\begin{array}{ccc}
0 & \Omega_{12}(t) & 0\\
\Omega_{12}(t) & 0 & \Omega_{23} (t)\\
0 & \Omega_{23}(t) & 0
\end{array}\right),
\end{eqnarray}
where all functions are assumed to be real. The typical counter-intuitive ordering of a STIRAP transfer is given by
\begin{eqnarray}
\Omega_{12} &=& \Omega_0 g \left[t - T (1/2 + \tau)\right],\\
\Omega_{23} &=& \Omega_0 g \left[t - T (1/2 - \tau)\right],
\end{eqnarray}
where $g(t) = \exp\left[-(t/T)^2/0.02\right]$ and the pulses are shown in Fig. \ref{fig_stirap_pulse}. The goal is to follow the usual dark state $\ket{\phi_{2}^{(0)}}$ which has eigenvalue $0$ always. Hence we define $\ket{\psi_{ad}(t)}=\ket{\phi_{2}^{(0)}(t)}$.

A Gaussian distribution is assumed for the strike strength of the noise $P(\xi) = \frac{1}{\sqrt{2\pi} \sigma} \fexp{-\frac{\xi^2}{2 \sigma^2}}$
with mean $\la\xi \ra = 0$ and width $\sigma$. The eigenvalues of $\cL_1$ can be found from the characteristic function of $P(\xi)$, namely
\begin{eqnarray}
\beta_{n,m} = \nu \left \{\exp \left[ -\frac{\sigma^2}{2} \left(\frac{E_n-E_m}{\hbar}\right)^2\right]-1 \right \}.
\end{eqnarray}
To numerically solve this, the master equation is represented in the eigenbasis of $\cL_1$.

%%%%%%%%%%%%%%%%%%%%%%%%%%%%%%%%%%%%%%%%%%
%Figure
%%%%%%%%%%%%%%%%%%%%%%%%%%%%%%%%%%%%%%%%%%
\begin{figure}[t]
\begin{center}
(a) \includegraphics[angle=0,width=0.7\linewidth]{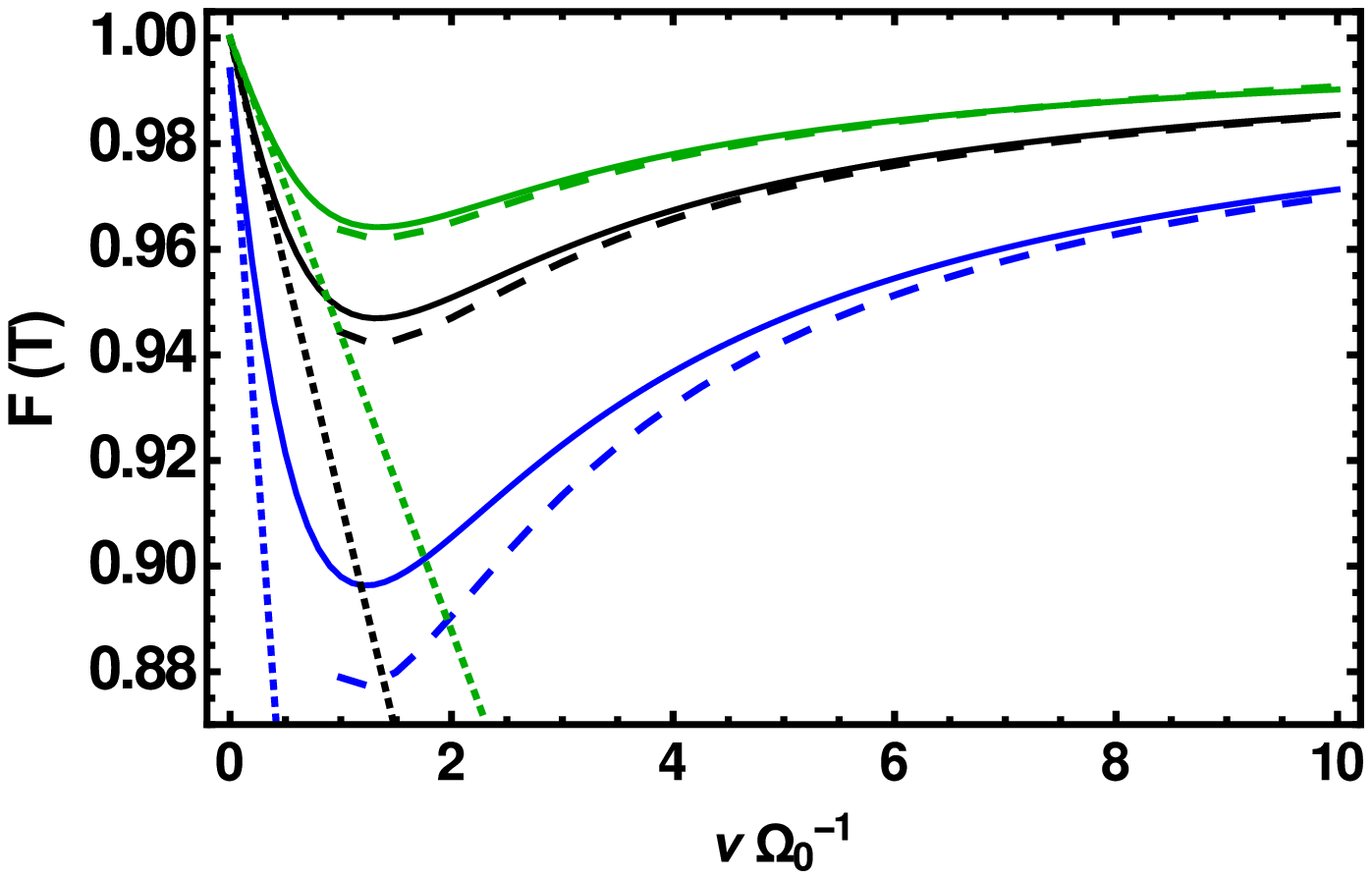} \\[0.5cm]%\hspace{0.5cm}
(b) \includegraphics[angle=0,width=0.7\linewidth]{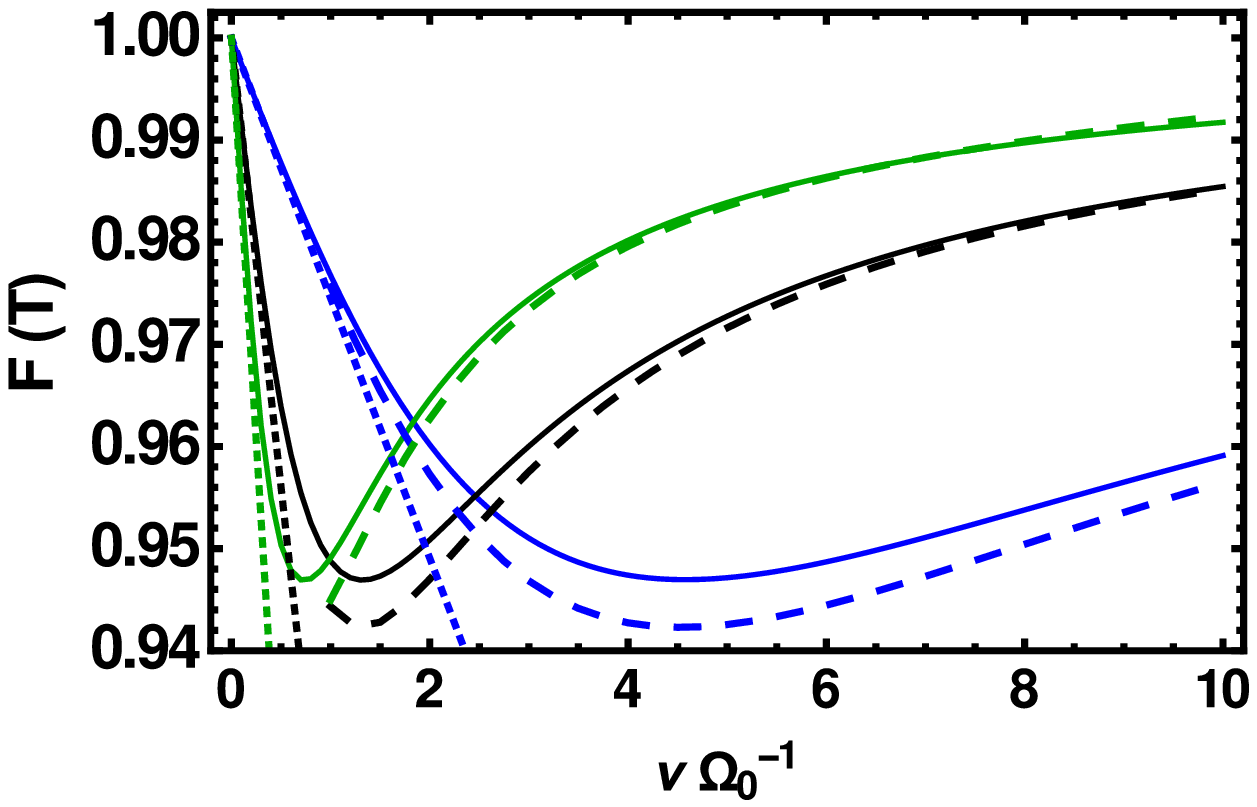}%\\[0.5cm]
\end{center}
\caption{STIRAP population transfer in a three-level system with $H_0 = H_1$, $\tau \Omega_{0}=0.1 $. Numerically exact solution (solid lines), small noise approximation Eq. \eqref{noiseapprox} (dotted lines) and strong noise approximation Eq. \eqref{secondorderapprox} (dashed line)
(a) Fidelity $F$ versus frequency $\nu$ for $\sigma \Omega_{0}=2 $ and $T \Omega_{0} =100,200,300$ blue, black and green respectively.
(b) Fidelity $F$ versus frequency $\nu$ for $T \Omega_{0} = 200$ and $\sigma \Omega_{0} =1, 2, 3$ blue, black and green respectively.}
\label{fig_stirap_1}
\end{figure}
%%%%%%%%%%%%%%%%%%%%%%%%%%%%%%%%%%%%%%%%%%
%Figure
%%%%%%%%%%%%%%%%%%%%%%%%%%%%%%%%%%%%%%%%%%

%%%%%%%%%%%%%%%%%%%%%%%%%%%%%%%%%%%%%%%%%%
%Figure
%%%%%%%%%%%%%%%%%%%%%%%%%%%%%%%%%%%%%%%%%%
\begin{figure}[t]
\begin{center}
(a) \includegraphics[angle=0,width=0.9\linewidth]{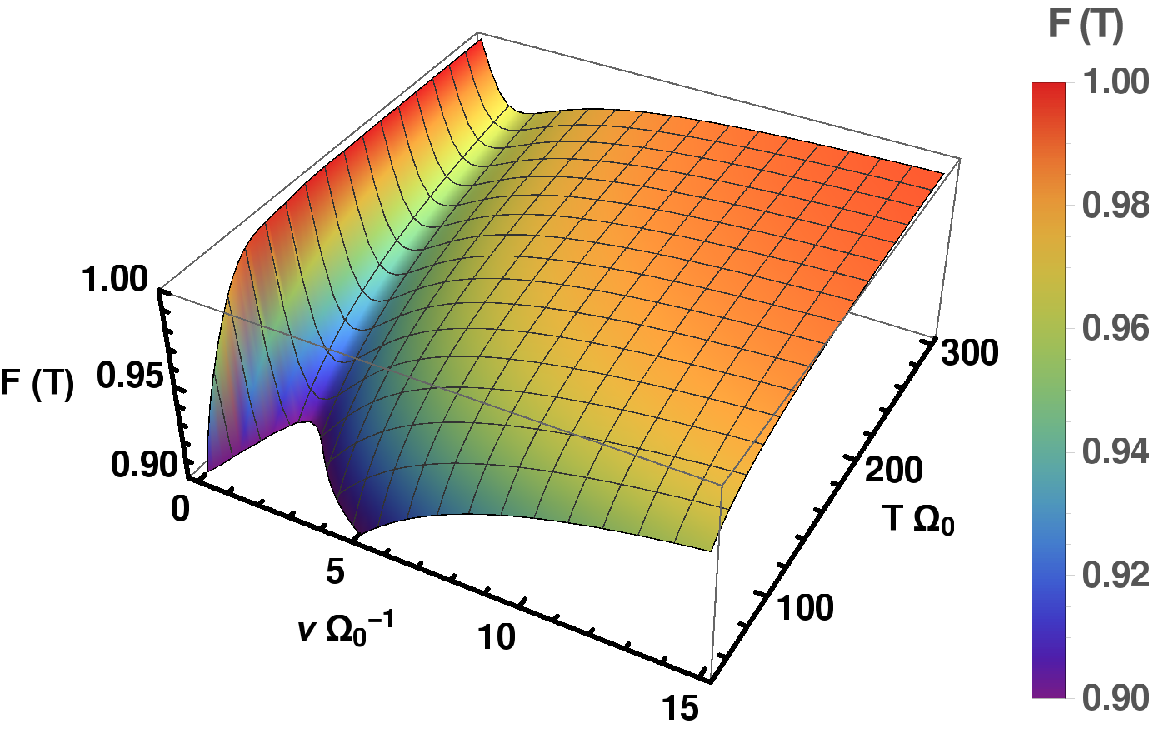}\\[0.5cm]
(b) \includegraphics[angle=0,width=0.9\linewidth]{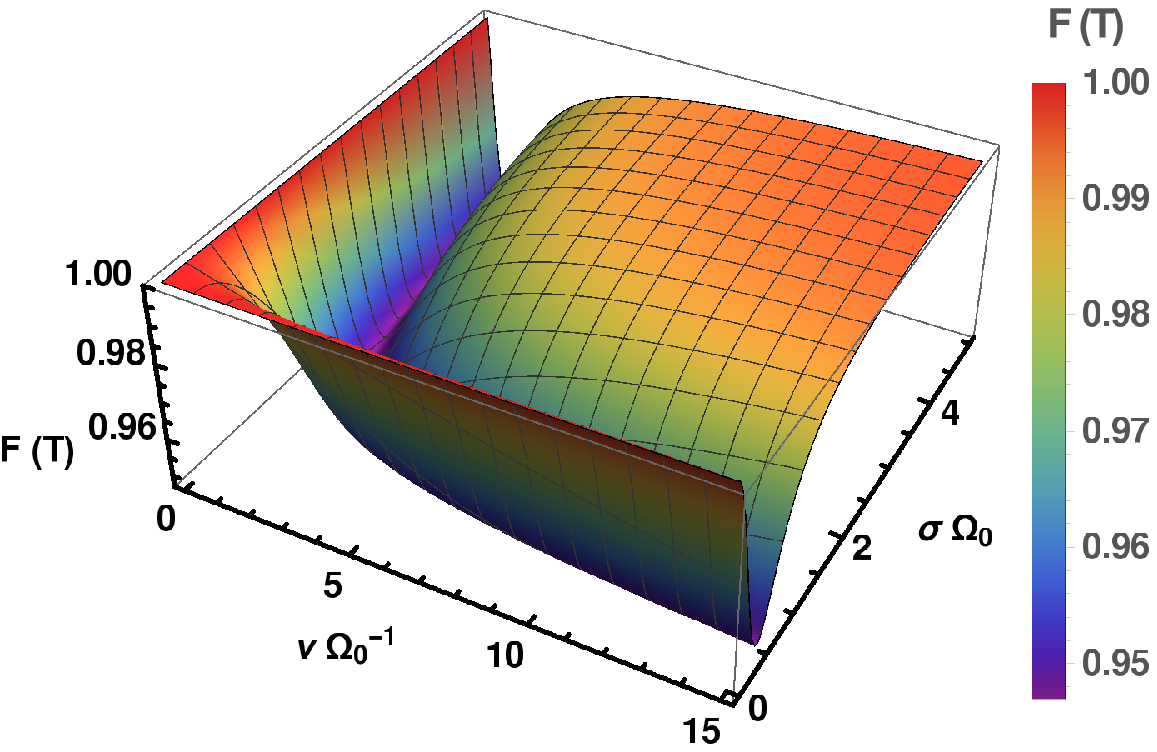}\\[0.5cm]
(c) \includegraphics[angle=0,width=0.9\linewidth]{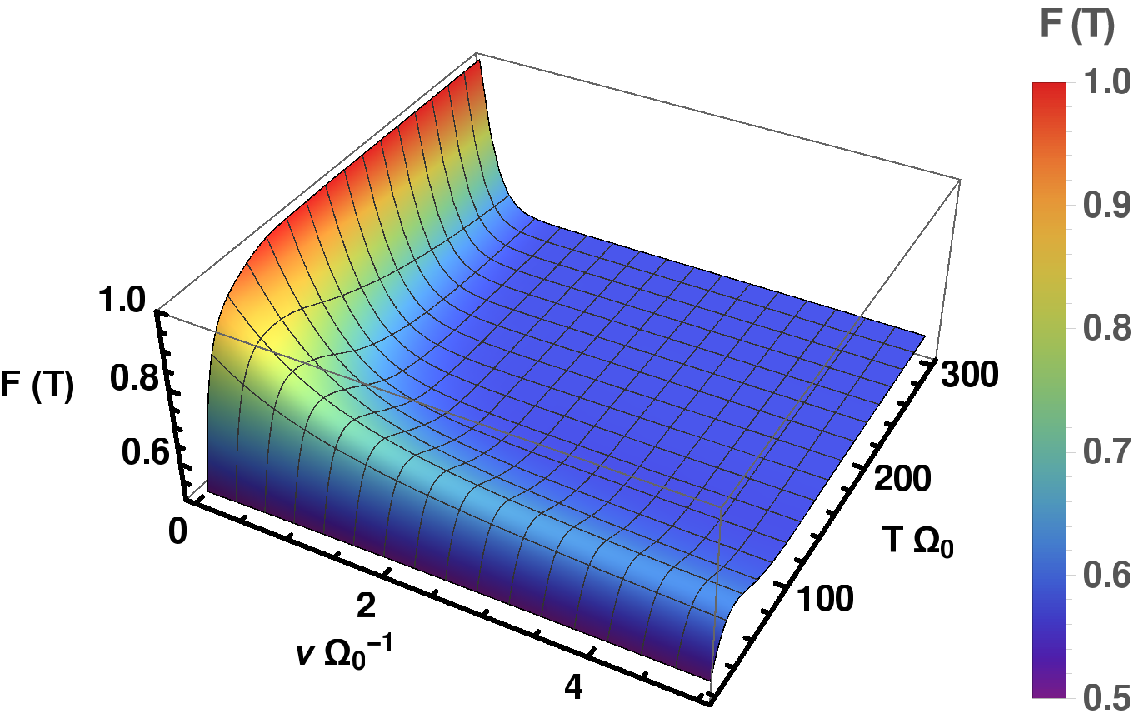}%\\[0.5cm]

\end{center}
\caption{STIRAP population transfer in a three-level system.
(a) Fidelity $F$ versus noise parameter $\nu$ and final time $T$ for $H_1 = H_0$; $\sigma \Omega_0 = 2$.
(b) Fidelity $F$ versus noise parameter $\nu$ and distribution width $\sigma$ for $H_1 = H_0$; $T \Omega_0=200$.
(c) Fidelity $F$ versus noise parameter $\nu$ and final time $T$ for the case of phase fluctuations; $\sigma \Omega_0 = 2$. $\tau \Omega_{0}=0.1 $ in all cases}
\label{fig_3}
\end{figure}
%%%%%%%%%%%%%%%%%%%%%%%%%%%%%%%%%%%%%%%%%%%
%Figure
%%%%%%%%%%%%%%%%%%%%%%%%%%%%%%%%%%%%%%%%%%

Different settings of the noise Hamiltonian are now considered.
In the first case, let $H_1 = H_0$. In Fig. \ref{fig_stirap_1}, the fidelity $F$ at final time $t=T$ against the frequency of the strikes $\nu$ is shown. The dip in the fidelity is present here again. In Fig. \ref{fig_stirap_1}(a) the fidelity is shown for different total times $T$. One can see that the amount that the fidelity drops is dictated by the adiabaticity (or equivalently the total time for the process). In Fig. \ref{fig_stirap_1}(b) the fidelity is shown for different distribution widths $\sigma$. The location of the turning point is determined by $\sigma$. In all cases the fidelity in the strong noise limit is $F_{\infty}=1$. The strong noise approximation (Eq. \eqref{secondorderapprox}) is seen to represent accurately the dynamics in the strong noise regime $\nu \Omega_{0}^{-1} \geq 1$.

In Fig. \ref{fig_3}(a) and (b) the fidelity is shown versus the final time $T$, the frequency of strikes $\nu$ and the variation in their strength $\sigma$. We can see again a dip which comes from the fact that small noise disturbs the adiabaticity while strong noise acts as a projector on the eigenstates of the noise Hamiltonian.

As an example where there is only decay in the fidelity consider the noise Hamiltonian
\begin{eqnarray}
H_1 &=& \frac{\hbar}{2} \left(\begin{array}{ccc}
0 & 0 & 0\\
0 & 0 & i\Omega_{23} (t)\\
0 & -i\Omega_{23}(t) & 0
\end{array}\right).\;
\end{eqnarray}
This could arise from random fluctuations in the phase of the Rabi frequency, i.e., $\Omega_{23}e^{i \kappa z(t)}\approx \Omega_{23} \left(1+i \kappa z(t)\right)$ for $\kappa \ll 1$. The fidelity in this case, plotted in Fig. \ref{fig_3}(c), goes down to a fixed value for increasing noise strength.

\section{Conclusion \label{Conclusion}}

Let us now summarize the work of the paper. We have presented a master equation for Poisson noise
in a general time dependent quantum system and outlined various properties associated with it.
We have outlined the behaviour in three regimes, namely adiabatic processes with no noise, weak noise
and strong noise. We have also shown that previous claims in \cite{garcia2014}, that white shot noise
can improve the adiabatic condition, may be misleading. Standard adiabaticity only improves when the noise bias is increased, i.e., when the Hamiltonian is made stronger, without necessarily implying a strong noise.  For very strong noise a different type of adiabaticity (in operator space rather than in the usual state space) emerges which implies the decay of coherences.  
Finally we have provided some numerical examples where this master equation can be used
for non-trivial systems such as a three-level system (where Poisson noise differs from standard Gaussian noise). In some examples, a dip in the fidelity as a function of noise strength is present where high fidelity still occurs for large noise strengths. Our results may also be relevant to describe continuous measurements, which are described by master equations which are formally similar
to those describing decoherence \cite{mensky,mensky_book1,mensky_book2}.

\section*{Acknowledgements}
We are grateful to David Rea for commenting on the manuscript and Lluc Garcia for useful discussions. This work was supported by MINECO (Grant No. FIS2015-67161-P); and the program UFI 11/55.

%----------------------------------------
\begin{appendix}
\section{Commutator recursion relation in two-level systems\label{app}}

Let us first define $\lambda=-\left(i/\hbar\right)\xi$ and then split the sum into even and odd terms,
\begin{eqnarray}
e^{\lambda H_{1}}\rho e^{-\lambda H_{1}} & = & \rho+\sum_{n=1}^{\infty}\frac{\lambda^{n}}{n!} \left[H_{1},\rho\right]_{n}\nonumber \\
 & = & \rho+\sum_{k=1}^{\infty}\frac{\lambda^{2k}}{\left(2k\right)!}\left[H_{1},\rho\right]_{2k} \nonumber \\ &+&\sum_{l=0}^{\infty}\frac{\lambda^{2l+1}}{\left(2l+1\right)!}\left[H_{1},\rho\right]_{2l+1} \nonumber \\
 & = & \rho+\sum_{k=1}^{\infty}\frac{\lambda^{2k}}{\left(2k\right)!}\chi^{k-1}2^{2\left(k-1\right)}\left[H_{1},\left[H_{1},\rho\right]\right] \nonumber \\ &+&\sum_{l=0}^{\infty}\frac{\lambda^{2l+1}}{\left(2l+1\right)!}\chi^{l}2^{2l}\left[H_{1},\rho\right] \nonumber \\
 & = & \rho+\left[\frac{-\sin^{2}\left(-\frac{1}{\hbar}\xi\sqrt{\chi}\right)}{2\chi}\right]\left[H_{1},\left[H_{1},\rho\right]\right] \nonumber \\
 &+& i\left[\frac{\sin\left(-\frac{2}{\hbar}\xi\sqrt{\chi}\right)}{2\sqrt{\chi}}\right]\left[H_{1},\rho\right].
\end{eqnarray}
We now need to prove the second last step. Let's do each case
separately. For $n$ odd it can be proved by induction that
\begin{eqnarray}
\left[H_{1},\rho\right]_{n}=2^{n-1}\chi^{\left(n-1\right)/2}\left[H_{1},\rho\right].
\end{eqnarray}
Clearly this is true for the case
of $n=1$. It can also be shown by explicit calculation to be true for $n=3$. Now let us show that if it is true for $n$ it is true
for $n+2$,
\begin{eqnarray}
\left[H_{1},\rho\right]_{n+2} & = & \left[H_{1},\left[H_{1},\left[H_{1},\rho\right]_{n}\right]\right] \nonumber \\
 & = & 2^{n-1}\chi^{\left(n-1\right)/2}\left[H_{1},\rho\right]_{3}\nonumber \\
 & = & 2^{n-1}\chi^{\left(n-1\right)/2}4\chi\left[H_{1},\rho\right] \nonumber \\
 & = & 2^{\left(n+2\right)-1}\chi^{\left(\left(n+2\right)-1\right)/2}\left[H_{1},\rho\right].
\end{eqnarray}
Hence it is true for all $n$ odd.

For $n$ even we claim that 
\begin{eqnarray}
\left[H_{1},\rho\right]_{n}=2^{n-2}\chi^{\left(n-2\right)/2}\left[H_{1},\left[H_{1},\rho\right]\right].
\end{eqnarray}
For $n=2$ and $n=4$ this holds true. Now let us show that if it
is true for $n$ it is true for $n+2$,
\begin{eqnarray}
\left[H_{1},\rho\right]_{n+2} & = & \left[H_{1},\left[H_{1},\left[H_{1},\rho\right]_{n}\right]\right] \nonumber \\
 & = & 2^{n-2}\chi^{\left(n-2\right)/2}\left[H_{1},\rho\right]_{4} \nonumber \\
 & = & 2^{\left(n+2\right)-2}\chi^{\left(\left(n+2\right)-2\right)/2}\left[H_{1},\left[H_{1},\rho\right]\right]. \nonumber \\
\end{eqnarray}
Hence it is true for all $n$ even.

\section{Derivation of strong noise limit \label{app2}}

In this section a more detailed overview of the derivation of the strong noise limit will be presented. Let us start by integrating Eq. \eqref{ceqstrong},
\begin{eqnarray}
c_{n,m}(T)&=&c_{n,m}(0)\nonumber \\ &+&\sum_{\substack{l,k \\ (l,k)\neq(n,m)}} \int^{T}_{0} dt \exp \left [ \tilde{\Lambda}_{l,k}(t)-\tilde{\Lambda}_{n,m}(t) \right] \nonumber \\ &\times &  M_{n,m,l,k}(t) c_{l,k}(t).\label{inteq}  
\end{eqnarray}
We then change to the coefficients
\begin{eqnarray}
d_{n,m}(t)=c_{n,m}(t) \exp \left[\tilde{\Lambda}_{n,m}(t)\right],
\end{eqnarray}
where $d_{n,m}(0)=c_{n,m}(0)$ since $\tilde{\Lambda}_{n,m}(0)=0$. By now rewriting Eq. \eqref{inteq} with these coefficients it becomes
\begin{eqnarray}
\lefteqn{d_{n,m}(T)=} && \nonumber\\
&& d_{n,m}(0)\exp\left[\tilde{\Lambda}_{n,m}(T)\right] \nonumber \\
&+& \sum_{\substack{l,k \\ (l,k)\neq(n,m)}} \int^{T}_{0} dt \exp \left [ \tilde{\Lambda}_{n,m}(T,t)\right] M_{n,m,l,k}(t) d_{l,k}(t), \nonumber \\ \label{deqapp}
\end{eqnarray}
where
\begin{eqnarray}
&&\tilde{\Lambda}_{n,m}(T,t)=\tilde{\Lambda}_{n,m}(T)-\tilde{\Lambda}_{n,m}(t) \nonumber \\
&=& \int_t^T ds\, \left[\kappa \beta_{n,m}(s) + \Sbraket{B_{n,m}(s)}{\cL_0|B_{n,m}(s)}\right]. \nonumber \\
\end{eqnarray}
This has the property that $\mbox{Re}\left(\tilde{\Lambda}_{n,m}(T,t)\right) \leq 0$ for $T \geq t$ and $\tilde{\Lambda}_{n,n}(T,t) = 0$. By using Eq. \eqref{deqapp} recursively one obtains the approximation
\begin{widetext}
\begin{eqnarray}
d_{n,m}(T)&\approx&  d_{n,m}(0)\exp\left[\tilde{\Lambda}_{n,m}(T)\right]+\sum_{\substack{l,k \\ (l,k)\neq(n,m)}} \int^{T}_{0} dt \exp \left [ \tilde{\Lambda}_{n,m}(T,t)+\tilde{\Lambda}_{l,k}(t) \right] M_{n,m,l,k}(t) d_{l,k}(0) \nonumber \\ &+&\sum_{\substack{l,k \\ (l,k)\neq(n,m)}} \sum_{\substack{q,r \\ (q,r)\neq(l,k)}} \int^{T}_{0} dt \int^{t}_{0} ds M_{n,m,l,k}(t) \exp \left [ \tilde{\Lambda}_{n,m}(T,t)+\tilde{\Lambda}_{l,k}(t,s)+\tilde{\Lambda}_{q,r}(s)\right] M_{l,k,q,r}(s) d_{q,r}(0),\nonumber \\ \label{secondorderapprox}
\end{eqnarray}
\end{widetext}
where the real part of all terms in the exponentials are negative. One can of course continue this process to obtain a series expansion on the right hand side. However for our purposes it is enough to understand the general form of the expansion so further terms are neglected.

For times $t_{2} > t_{1}$ it is clear that $\exp\left[\tilde{\Lambda}_{n,m}(t_{2},t_{1})\right] \rightarrow 0$ as $\kappa \rightarrow \infty$ if $n \neq m$. However for $n=m$, $\exp\left[\tilde{\Lambda}_{n,n}(t_{2},t_{1})\right]=1$ for all $\kappa$. Recall that we assume $\beta_{n,m}=0$ if and only if $n=m$. It is then straightforward to see that $d_{n,m}(T) \rightarrow 0$ as $\kappa \rightarrow \infty$ for $n \neq m$. An approximation to this is $d_{n,m}(T) \approx d_{n,m}(0) \exp\left[\tilde{\Lambda}_{n,m}(T)\right]$. Converting back to the original coefficients we get that $c_{n,m}(T) \approx c_{n,m}(0)$ for large $\kappa$.

The result is more difficult to see if $n=m$.
By explicit calculation it is found that
\begin{eqnarray}
\lefteqn{M_{n,n,l,k}(t)=}&& \nonumber\\
&& \begin{cases} 
\frac{i}{\hbar}\bra{\phi^{(1)}_{k}}H_{0}\ket{\phi^{(1)}_{n}}-\braket{\dot{\phi}^{(1)}_{k}}{\phi^{(1)}_{n}} & l = n,k \neq n \\
-\frac{i}{\hbar}\bra{\phi^{(1)}_{n}}H_{0}\ket{\phi^{(1)}_{l}}-\braket{\phi^{(1)}_{n}}{\dot{\phi}^{(1)}_{l}} & l \neq n,k = n \\
0 & l \neq n, k \neq n \\
0 & l =n, k=n
\end{cases}. \nonumber \\
\end{eqnarray}
Therefore all cases where $M_{n,n,l,k}(t) \neq 0$ have $l \neq k$. However in this case $\exp \left[\tilde{\Lambda}_{l,k}(t,0)\right] \rightarrow 0$ in the limit where $\kappa \rightarrow \infty$. In terms of the original coefficients this gives $c_{n,n}(T) \approx c_{n,n}(0)$ for large $\kappa$. So in general we get that $c_{n,m}(T) \approx c_{n,m}(0)$ for all $n,m$ in the case of strong noise i.e. large $\kappa$.

\end{appendix}

%\section*{References}

\end{document}